\documentclass[prl,twocolumn,showpacs,amsmath,amssymb]{revtex4}
\usepackage{graphicx}% Include figure files
\usepackage{dcolumn}% Align table columns on decimal point
\usepackage{bm}% bold math

\begin{document}

\title{Chiral spin pairing in helical magnets}

\author{Shigeki Onoda}
\altaffiliation{
Present address: RIKEN (The Institute of Physical and Chemical Research), Hirosawa, Wako 351-0198, Japan}
\email{s.onoda@riken.jp}
\affiliation{%
CREST, Department of Applied Physics, University of Tokyo, Tokyo 113-8656, Japan}
\author{Naoto Nagaosa}%
\affiliation{%
CREST, Department of Applied Physics, University of Tokyo, Tokyo 113-8656, Japan}
\date{\today}% It is always \today, today,

\begin{abstract}
  A concept of chiral spin pairing is introduced to describe a vector-chiral liquid-crystal order in frustrated spin systems. It is found that the chiral spin pairing is induced by the coupling to phonons through the Dzyaloshinskii-Moriya interaction and the four-spin exchange interaction of the Coulomb origin under the edge-sharing network of magnetic and ligand ions. This produces two successive second-order phase transitions upon cooling: an O(2) chiral spin nematic, i.e., spin cholesteric, order appears with an either parity, and then the O(2) symmetry is broken to yield a helical magnetic order. Possible candidate materials are also discussed as new multiferroic systems.
\end{abstract}

\maketitle

The chirality in the electronic spin texture introduced by a geometrical frustration and/or the relativistic spin-orbit interaction has been one of the key concepts in strongly correlated electron systems. The scalar spin chirality~\cite{BaskaranAnderson88,Laughlin88,WenWilczekZee89}, i.e., the scalar triple product of three noncoplanar spins, is odd under the time-reversal ($T$) and even under the space-inversion ($I$), and gives rise to a large anomalous Hall effect~\cite{Ong98,Ohgushi00,Taguchi01,Onoda03,Machida07}. On the other hand, central to this Letter is the vector spin chirality~\cite{Villain}, i.e., the vector product of two noncollinear spins. It is $T$-even and $I$-odd, and can produce the ferroelectric polarization in Mott insulators through the spin-orbit interaction~\cite{Kimura03,Katsura05,Kenzelman05,Lawes05,Arima06,Mostovoy06,Harris06,JONH07}, even for spin-1/2 systems~\cite{JONH07,LiCuVO2,LiCu2O2}.

The vector-chiral spin order is realized in conventional helical magnets. In principle, it is even possible that the chiral or parity symmetry is broken but the time-reversal symmetry is not. This state having finite spin correlation lengths is categorized into a liquid or a liquid crystal of spins, which is of our main interest, while the magnetically ordered state into a solid. The issue of the chiral spin order in the absence of any magnetic order is not only a problem of statistical mechanics, but has been intensively studied since the discovery of high-$T_c$ cuprates, and even gives a new ``multiferroic'' phenomenon which has attracted current great interests, as discussed below. However, conditions for the order being stabilized have not been fully understood.

It has been argued that the vector or pseudoscalar chiral ordering~\cite{Teitel83,Andreev,MiyashitaShiba84} takes place separately from the XY transition in the antiferromagnetic (AF) XY model on the triangular lattice~\cite{Lee98}. The intermediate phase is characterized by the vector-chiral order. It has also been obtained for frustrated quantum spin systems in one dimension~\cite{Hikihara}, and discussed in two dimensions~\cite{Chandra90} and for odd-time states~\cite{Balatsky}. However, it usually appears only in a tiny region of the global phase diagram. For classical AF Heisenberg and XY models on stacked triangular lattices, Monte-Carlo simulations and field-theoretical analyses have suggested a single phase transition from paramagnet to helical magnet, which is either weakly first-order or a second-order one, which belongs to a different universality class from the O(N) model~\cite{Kawamura_rev,Calabrese01,Holovatch04}. Then, the chiral spin liquid-crystal phase does not appear.

\begin{figure}[htb]
\begin{center}
  \includegraphics[width=7.8cm]{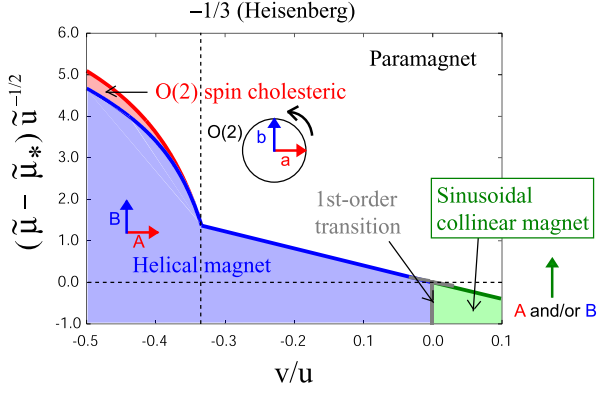}
\end{center}
\caption{(Color online) Phase diagram of the Ginzburg-Landau Hamiltonian given by Eq.~(\ref{eq:H_r}) and the ordering patterns. Note that $\tilde{\mu}_*=-8\tilde{u}$. For details, see the text.}
\label{fig:phase}
\end{figure}

In this Letter, we introduce a picture of chiral spin pairing for the vector-chiral spin order which has an analogy to a cholesteric liquid crystal defined as a chiral nematic~\cite{degenne}, and propose a new strategy to realize this phase. It is found that when the coupling of spins to optical phonons through the Dzyaloshinskii-Moriya (DM) interaction and/or the Coulombic four-spin ring-exchange interaction under the edge-sharing network of magentic and ligand ions are robust against the magnetostriction, these interactions bind spin pairs with the vector chirality without any magnetic long-range order. Then, upon cooling, the transition from paramagnet to helical magnet takes place through two steps of lowering the symmetry. Namely, an O(2) spin cholesteric and a helical magnetic phases succesively appear. (See Fig.~\ref{fig:phase}).

We begin with the Ginzbrug-Landau Hamiltonian for incommensurate helical Heisenberg magnets with the ordering wave vector $\bm{Q}$, which can be expressed in terms of two independent long wavelength modes of the O(3) field $\vec{S}_{\bm{r}}$ for the spin at a site $\bm{r}$, namely, $\vec{S}_{\bm{q}}=\sum_{\bm{r}}\vec{S}_{\bm{r}}e^{-i\bm{q}\cdot\bm{r}}$ with $\bm{q}\sim\pm\bm{Q}$, or equivalently the long-wavelength modes
$\vec{a}_{\bm{q}}\equiv\frac{1}{2}(\vec{S}_{\bm{Q}+\bm{q}}+\vec{S}_{-\bm{Q}+\bm{q}})$ and $\vec{b}_{\bm{q}}\equiv-\frac{i}{2}(\vec{S}_{\bm{Q}+\bm{q}}-\vec{S}_{-\bm{Q}+\bm{q}})$
with $|\bm{q}|<\Lambda$ and the momentum cutoff $\Lambda$. This gives
\begin{equation}
  \vec{S}_{\bm{r}}\approx\vec{a}_{\bm{r}}\cos\bm{Q}\cdot\bm{r}-\vec{b}_{\bm{r}}\sin\bm{Q}\cdot\bm{r},
  \label{eq:S_r:ab}
\end{equation}
in terms of the inverse Fourier transforms
$(\vec{a}_{\bm{r}},\vec{b}_{\bm{r}})
=a_0^d\int^\Lambda\!\!\frac{d^d\bm{q}}{(2\pi)^d}\,
(\vec{a}_{\bm{q}},\vec{b}_{\bm{q}}) e^{i\bm{q}\cdot\bm{r}}$
with the lattice constant $a_0$. Here, ``$\approx$'' represents a weak equality except the irrelevant rapidly varying fields with $|\bm{q}|>\Lambda$. Substituting Eq.~(\ref{eq:S_r:ab}) into the Heisenberg Hamiltonian, one obtains the continuum model~\cite{Kawamura_rev} given by
\begin{eqnarray}
  {\cal H}&\approx&\int\!d^d\bm{r}\Bigl[\frac{\mu}{2}(\vec{a}^2+\vec{b}^2)+\frac{1}{2}\left((\nabla\vec{a})^2+(\nabla\vec{b})^2\right)
    \nonumber\\
    &&\ \ \ \ \ \ \ +\frac{u}{4}(\vec{a}^2+\vec{b}^2)^2
    +v((\vec{a}^2\vec{b}^2)-(\vec{a}\cdot\vec{b})^2)\Bigr].
  \label{eq:H_r}
\end{eqnarray}
We have neglected the spatial anisotropy in the gradient terms for simplicity, and rescaled the variables so that coefficients to those terms become unity.
Quartic interaction terms with the coupling constants $u=u_0>0$ and $v=v_0=-u_0/3$ can be obtained by softening the constraint that amplitudes of spins must be fixed, $\vec{S}(\bm{r})^2\approx S^2$, and substituting Eq.~(\ref{eq:S_r:ab}) into the quartic term $\frac{u_0}{6}\int\!d\bm{r}(\vec{S}_{\bm{r}}^2)^2$.
Then, the Hamiltonian given by Eq.~(\ref{eq:H_r}) is isomorphic to a classical O(3)$\times$O(2) nonlinear-$\sigma$ model~\cite{Kawamura_rev}. 
Now helical and collinear magnetic orders are expressed by $\langle\vec{a}\rangle\times\langle\vec{b}\rangle\ne\vec{0}$ and $\langle\vec{a}\rangle\times\langle\vec{b}\rangle=\vec{0}$, respectively. The vector-chiral order is characterized by $\langle \vec{a}\times\vec{b}\rangle\ne\vec{0}$. $u$ and $v$ are modified from the Heisenberg values by additional interactions as shown later. 

Figure~\ref{fig:phase} presents the main result of this paper, i.e., the phase diagram for $d=3$ obtained in the space of $(\tilde{\mu},v/u)$ for the model Eq.~(\ref{eq:H_r}). Here we introduce dimensionless coupling constants $\tilde{u}\equiv ua_0^3/(2\pi^2\Lambda)$ and $\tilde{v}\equiv va_0^3/(2\pi^2\Lambda)$, as well as $\tilde{\mu}\equiv\mu/\Lambda^2$ which corresponds to the temperature. The spin cholesteric, i.e., chiral spin pairing state, appears above the helical magnetic ordering transition temperature when $v/u$ is less than -1/3. When $\tilde{\mu}$ is negatively large, a negative/positive $v$ stabilizes a helical/collinear magnetic order. These two phases are separated by the first-order phase transition at $v=0$. 

To derive this result, we have employed the mode coupling approximation, where the average value and the optimal Gaussian fluctuation form of the order parameters are determined by the variational principle~\cite{Moriya}. 
First, we decompose the fields into the condensed and normal components as
$(\vec{a}_{\bm{q}},\vec{b}_{\bm{q}})=\delta_{\bm{q}}(\vec{A},\vec{B})+(\delta\vec{a}_{\bm{q}},\delta\vec{b}_{\bm{q}})$.
$\langle\vec{a}_{\bm{r}}\rangle=\vec{A}$ and $\langle\vec{b}_{\bm{r}}\rangle=\vec{B}$ give the ordered mangetic moment through Eq.~(\ref{eq:S_r:ab}).
$(\vec{A},\vec{B})$ is subject to the stationary condition of the free energy with respect to $(\delta\vec{a},\delta\vec{b})$.
To treat the Gaussian fluctuations around the saddle point, we consider the variational quadratic Hamiltonian 
${\cal H}_{\text{var}}=(a_0/2\pi)^{-3}\int\!d\bm{q}\sum_{i,j=1}^3 \ 
(\delta a^i_{\bm{q}},\delta b^i_{\bm{q}}) \ 
\hat{G}^{-1}_{\bm{q}}{}^{ij} \ 
{}^t(\delta a^j_{-\bm{q}},\delta b^j_{-\bm{q}})$ with the $6\times6$-matrix Green's function $G_{\bm{q}}$ defined by
\begin{equation}
  \hat{G}^{-1}_{\bm{q}}{}^{ij}=(\delta_{i,j}\bm{q}^2+\mu^{ij})\hat{\rho}^0+\delta\mu^{ij}\hat{\rho}^z+\Delta_s^{ij}\hat{\rho}^x+\Delta_a^{ij}i\hat{\rho}^y
  \label{eq:G}
\end{equation}
with the $2\times2$ identity matrix $\hat{\rho}^0$ and the Pauli matrices $(\hat{\rho}^x,\hat{\rho}^y,\hat{\rho}^z)$ operating on the $(a,b)$ space. Here, the variational parameters $\mu^{ij}$, $\delta\mu^{ij}$, $\Delta_s^{ij}$, and $\Delta_a^{ij}$ are determined from the self-consistent equations,
\begin{subequations}
\begin{eqnarray}
  \mu^{ij}&=&\delta^{ij}\Bigl[\mu+(u+v)\sum _\ell\left({A^\ell}^2+{B^\ell}^2+\gamma_0^{\ell\ell}\right)\Bigr]
  \nonumber\\
  &&\ \ \ \ \ \ \ \ {}+(u-v)\left(A^iA^j+B^iB^j+\gamma_0^{ij}\right),\ \ \ 
  \label{eq:mu}\\
  \delta\mu^{ij}&=&-v\delta^{ij}\sum_\ell\left({A^\ell}^2-{B^\ell}^2+\gamma_z^{\ell\ell}\right)
  \nonumber\\
  &&{}+(u+v)\left(A^iA^j-B^iB^j+\gamma_z^{ij}\right),
  \label{eq:dmu}\\
  \Delta^{ij}_s&=&-v\delta_{i,j}\sum_\ell\left(2A^\ell B^\ell+\gamma_x^{\ell\ell}\right)
  \nonumber\\
  &&+(u+v)\left(A^iB^j+A^jB^i+\gamma_x^{ij}\right),
  \label{eq:Delta_s}\\
  \Delta^{ij}_a&=&(u+3v)\left(A^iB^j-A^jB^i+i\gamma_y^{ij}\right),
  \label{eq:Delta_a}
\end{eqnarray}
\label{eq:sc}
\end{subequations}
where 
$\gamma_\mu^{ij}\equiv\langle
 \delta a_{\bm{r}}^i \delta a_{\bm{r}}^j
+\delta b_{\bm{r}}^i \delta b_{\bm{r}}^j\rangle$,
$\gamma_x^{ij}\equiv\langle
 \delta a_{\bm{r}}^i \delta b_{\bm{r}}^j
+\delta b_{\bm{r}}^i \delta a_{\bm{r}}^j\rangle$,
$\gamma_z^{ij}\equiv\langle
 \delta a_{\bm{r}}^i \delta a_{\bm{r}}^j
-\delta b_{\bm{r}}^i \delta b_{\bm{r}}^j\rangle$,
$\gamma_x^{ij}\equiv\langle
 \delta a_{\bm{r}}^i \delta b_{\bm{r}}^j
+\delta b_{\bm{r}}^i \delta a_{\bm{r}}^j\rangle$, and
$i\gamma_y^{ij}\equiv\langle
 \delta a_{\bm{r}}^i \delta b_{\bm{r}}^j
-\delta b_{\bm{r}}^i \delta a_{\bm{r}}^j\rangle$ are calculated from
\begin{equation}
  \gamma_\mu^{ij}=a_0^3\!\!\int\!\!\frac{d\bm{q}}{(2\pi)^3}\,\text{Tr}\hat{G}^{ij}_{\bm{q}}\hat{\rho}^\mu,
  \label{eq:abiabj}
\end{equation}
where the trace Tr is taken only in the $(a,b)$ space.

We start from the disordered phase where all the variational parameters including $(\vec{A},\vec{B})$ vanish except the mass $\mu^r\equiv\mu^{ii}$. Then, the self-consistent equations~(\ref{eq:sc}) can be linearized with respect to the dimensionless variational order parameters $\tilde{\mu}^{ij}\equiv\mu^{ij}/\Lambda^2$ ($i\ne j$), $(\delta\tilde{\mu}^{ij},\tilde{\Delta}_s^{ij})\equiv(\delta\mu^{ij},\Delta_s^{ij})/\Lambda^2$, and $\tilde{\Delta}_a^{ij}\equiv\Delta_a^{ij}/\Lambda^2$;
\begin{subequations}
\begin{eqnarray}
  \tilde{\mu}^r&\equiv&\mu^r/\Lambda^2=\tilde{\mu}+4(2\tilde{u}+\tilde{v})g(\tilde{\mu}^r),
  \label{eq:mu^r:1}\\
  \tilde{\mu}^{ij}&=&-2(\tilde{u}-\tilde{v})\tilde{\mu}^{ij}f(\tilde{\mu}^r)
  \ \ \ \mbox{for $i\ne j$},\ \ \ \
  \label{eq:mu:1}\\
  \left(\begin{array}{c}
    \delta\tilde{\mu}^{ij}\\
    \tilde{\Delta}_s^{ij}
  \end{array}\right)&=&-2\left(\begin{array}{c}
    \delta\tilde{\mu}^{ij}\\
    \tilde{\Delta}_s^{ij}
  \end{array}\right)\left[-3\tilde{v}\delta_{ij}+(\tilde{u}+\tilde{v})\right]f(\tilde{\mu}^r),\ \ \ \ \ 
  \label{eq:dmu,Ds:1}\\
  \tilde{\Delta}_a^{ij}&=&-2\tilde{\Delta}_a^{ij}(\tilde{u}+3\tilde{v})f(\mu^r),
  \label{eq:Da:1}
\end{eqnarray}
\end{subequations}
with
\begin{subequations}
\begin{eqnarray}
  g(\tilde{\mu}^r)&=&1-\sqrt{\tilde{\mu}^r}\arctan\frac{1}{\sqrt{\tilde{\mu}^r}},
  \label{eq:g}\\
  f(\tilde{\mu}^r)&=&\frac{1}{2}\left[\frac{1}{\sqrt{\tilde{\mu}^r}}\arctan\frac{1}{\sqrt{\tilde{\mu}^r}}-\frac{1}{1+\tilde{\mu}^r}\right].
  \label{eq:f}
\end{eqnarray}
\end{subequations}
If we ignore the spin pairing, i.e., $\delta\mu^{ij}=\Delta_s^{ij}=\Delta_a^{ij}=0$, Eq.~(\ref{eq:mu^r:1}) always gives a direct phase transition from the paramagnet to a helical (collinear) magnet for $v<0$ ($v>0$) at a critical point $\tilde{\mu}^r=0$. However, a serious consideration on the spin pairs $\delta\mu^{ij}$, $\Delta_s^{ij}$, and $\Delta_a^{ij}$ reveals that a spin liquid-crystal phase emerges before the magnetic order sets in.
Note that $g(\tilde{\mu}^r)$ and $f(\tilde{\mu}^r)$ are positive and that $f(\mu^r)$ diverges as $\pi/(4\sqrt{\tilde{\mu}^r})$ for $\tilde{\mu}^r\to0$. Then, as far as the Hamiltonian is stable, i.e., $u+v>0$, there appear two possibilities for spin paired states. (i) When the interaction for the antisymmetric spin pairing channel is attractive, namely, $v/u=\tilde{v}/\tilde{u}<-1/3$ in Eq.~(\ref{eq:Da:1}), then it forms the chiral spin pairs $\Delta_a^{ij}\ne0$ through a second-order phase transition. The critical value of $\mu^r$ is determined by $f(\tilde{\mu}^r_{\mathrm{cr}})=-1/2(\tilde{u}+3\tilde{v})$ from Eq.~(\ref{eq:Da:1}) with  $\tilde{\mu}^r_{\mathrm{cr}}>0$: the magnetic long-range order is absent. This represents a O(2)-symmetric chiral spin nematic~\cite{Balatsky} or spin cholesteric. (ii) When $v/u=\tilde{v}/\tilde{u}>1/2$ in Eq.~(\ref{eq:dmu,Ds:1}), the symmetric spin pairs are formed, $(\delta\mu^{ii},\Delta_s^{ii})\ne0$. However, this condition is usually difficult to be realized starting from the Heisenberg point $v/u=-1/3$. In the following, we concentrate on the case of (i) $v/u<-1/3$, which is more likely to occur. 

In the spin cholesteric phase characterized by $\Delta_a^{ij},\gamma^{ij}_y\propto\epsilon_{ijz}$ with the fully antisymmetric tensor $\epsilon_{ij\ell}$, the Green's function can be described by
\begin{equation}
  \hat{G}^{ij}_{\bm{q}}=\frac{1}{2}\sum_{\sigma=\pm}\frac{\delta_{ij}-i\sigma\epsilon_{ijz}\hat{\rho}^y}{\mu^{ii}+\bm{q}^2-\sigma(1-\delta_{iz})\Delta_a}.
  \label{eq:Green}
\end{equation}
Here, $\mu_\parallel\equiv \mu^{xx}=\mu^{yy}$, $\mu_\perp\equiv\mu^{zz}$, and $\Delta_a$ are determined by the numerical solution to the self-consistent equations,
\begin{subequations}
\begin{eqnarray}
  \tilde{\mu}_\perp&=&\tilde{\mu}+2\tilde{u}h^+(\tilde{\mu}_\perp,0)+2(\tilde{u}+\tilde{v})h^+(\tilde{\mu}_\parallel,\tilde{\Delta}_a),
  \label{eq:gap:mu_out}\\
  \tilde{\mu}_\parallel&=&\tilde{\mu}+(\tilde{u}+\tilde{v})h^+(\tilde{\mu}_\perp,0)+(3\tilde{u}+\tilde{v})h^+(\tilde{\mu}_\parallel,\tilde{\Delta}_a),\ \ \ \ \ 
  \label{eq:gap:mu_in}\\
  \tilde{\Delta}_a&=& -(\tilde{u}+3\tilde{v})h^-(\tilde{\mu}_\parallel,\tilde{\Delta}_a),
  \label{eq:gap:cholesteric}
\end{eqnarray}
\end{subequations}
which are derived from Eqs.~(\ref{eq:mu}) and (\ref{eq:Delta_a}), where
$h^\pm(\tilde{\mu},\tilde{\Delta}_a)=g(\tilde{\mu}-\tilde{\Delta}_a)\pm g(\tilde{\mu}+\tilde{\Delta}_a)$.
Since $\mu_\parallel<\mu_\perp$ holds generally, the spins that are originally O(3)-symmetric now obtain the easy-plane XY anisotropy.
Accordingly, the spin correlation has three nonvanishing branches;
\begin{eqnarray}
  \langle S^z_{\bm{r}}S^z_{\bm{r}'}\rangle &=& \chi_{\bm{r},\bm{r}'}(\mu_\perp)\cos\bm{Q}\cdot(\bm{r}-\bm{r}'),
  \label{eq:S^zS^z}\\
  \langle S^i_{\bm{r}}S^i_{\bm{r}'}\rangle &=& \frac{1}{2}\sum_{\sigma=\pm}\chi_{\bm{r},\bm{r}'}(\xi_{\parallel\sigma}^{-2})\cos\bm{Q}\cdot(\bm{r}-\bm{r}'),\ \ \ \ \ 
  \label{eq:S^xyS^xy}
\end{eqnarray}
with $i=x,y$ and the chiral spin correlation
\begin{equation}
  \langle S^x_{\bm{r}}S^y_{\bm{r}'}-S^y_{\bm{r}}S^x_{\bm{r}'}\rangle
  =\sum_{\sigma=\pm}\sigma\chi_{\bm{r},\bm{r}'}(\xi_{\parallel\sigma}^{-2})\sin\bm{Q}\cdot(\bm{r}-\bm{r}'),\ \ 
  \label{eq:S^xS^y-S^yS^x}
\end{equation}
where $\xi_\sigma\equiv1/\sqrt{\mu_\parallel-\sigma \Delta_a}$ and
\begin{equation}
  \chi_{\bm{r},\bm{r}'}(\xi^{-2})
  \sim\frac{a^3\Lambda}{4\pi^2}\frac{\pi}{\Lambda|\bm{r}-\bm{r}'|}\exp\left(-\xi|\bm{r}-\bm{r}'|\right)
  \label{eq:chi^pm}
\end{equation}
for a long-distance decay.
These spin correlations are all short-ranged with two correlation lengths $\sim\xi_\pm$ . The sign of the chirality $\Delta_a$ can be determined from that of 
$\langle S^x_{\bm{r}}S^y_{\bm{r}'}-S^y_{\bm{r}}S^x_{\bm{r}'}\rangle$. 
Further decreasing $\mu$, $\tilde{\mu}_\parallel$ becomes equal to $\Delta_a$, leading to a divergence of either $\xi_+$ or $\xi_-$ and thus the second-order phase transition to the helical magnet.
Note that all the above structures should appear in the static spin correlations,
which can be directly studied by the polarization dependence of the incident and/or scattered neutrons~\cite{Maleyev}. 

In fact, the coupling constants $u$ and $v$ are modified by (i) the coupling of spins to longitudinal phonons with the wavevector $\pm2\bm{Q}$ through the magnetostriction, (ii) that to transverse phonons with the wavevector $\bm{0}$ through the DM interaction, and (iii) the four-spin ring-exchange interactions, i.e., $u=u_0+\delta u^{\text{ms}}+\delta u^{\text{DM}}+\delta u^{\text{ring}}$ and $v=-u_0/3+\delta v^{\text{ms}}+\delta v^{\text{DM}}+\delta v^{\text{ring}}$. 

(i) The magnetostriction arises from
\begin{equation}
  -\delta J\sum_{\bm{r},\bm{r}'}^{\text{n.n.}}(\vec{X}_{\bm{r}'}-\vec{X}_{\bm{r}})\cdot\vec{e}_{\bm{r}'-\bm{r}}\vec{S}_{\bm{r}}\cdot\vec{S}_{\bm{r}'}+\sum_{\bm{r}}\biggl[\frac{\vec{P}_{\bm{r}}^2}{2M}+\frac{1}{2}K\vec{X}_{\bm{r}}^2\biggr]
  \label{eq:H:MS}
\end{equation}
with the spatial derivative $\delta J$ of the exchange interaction, the unit vector $\vec{e}_{\bm{r}'-\bm{r}}$ directing from $\bm{r}$ to $\bm{r}'$, the mass $M$ of the magnetic ion, the spring constant $K$ associated with its shift $\vec{X}_{\bm{r}}$, and the conjugate momentum $\vec{P}_{\bm{r}}$. Substituting Eq.~(\ref{eq:S_r:ab}) into Eq.~(\ref{eq:H:MS}) and integrating over $\vec{X}_{\bm{r}}$, we obtain $\delta u^{\text{MS}}=\delta v^{\text{MS}}=(c^{-2}\delta J\sin\bm{Q}\cdot\bm{\delta})^2/4K$ where $c$ is the average spin-wave velocity.

(ii) Additional terms to the Hamiltonian reads
\begin{eqnarray}
  \frac{1}{2}\sum_{\bm{r},\bm{r}'}^{n.n.}\biggl[-\vec{D}_{\bm{r},\bm{r}}\cdot\vec{S}_{\bm{r}}\times\vec{S}_{\bm{r}'}
    +\frac{\vec{p}_{\bm{r},\bm{r}'}^2}{2m_L}+\frac{1}{2}\kappa\vec{x}_{\bm{r},\bm{r}'}^2\biggr].\ \ \ 
  \label{eq:H:DM}
\end{eqnarray}
Here, $\vec{D}_{\bm{r},\bm{r}'}=\lambda\vec{x}_{\bm{r},\bm{r}'}\times\vec{e}_{\bm{r}-\bm{r}'}$ represents the anti-symmetric DM coupling vector~\cite{DM} between the nearest-neighbor spin pair $\vec{S}_{\bm{r}}$ and $\vec{S}_{\bm{r}'}$ located at the discrete lattice sites $\bm{r}$ and $\bm{r}'$, respectively, with the coupling constant $\lambda$ and a transverse shift $\vec{x}_{\bm{r},\bm{r}'}$ of the ligand ion~\cite{JONH07}. For simplicity, we have assumed that the ligand ion is located at the center of nearest-neighbor bonds of spins. Because of this coupling, the chiral order $\langle \vec{a}\times\vec{b}\rangle\ne\vec{0}$ simultaneously produces both macroscopic averages of $\vec{x}$ and the electric polarization in proportion to $\langle\vec{a}\times\vec{b}\rangle$ and $\bm{Q}\times(\langle\vec{a}\times\vec{b}\rangle)$, respectively. Note that applied electric field plays the same role as the average value of $\vec{x}$. Substituting Eq.~(\ref{eq:S_r:ab}) into Eq.~(\ref{eq:H:DM}) and integrating over $\vec{x}_{\bm{r},\bm{r}'}$, we obtain $\delta u^{\text{DM}}=0$ and $\delta v^{\text{DM}}=-(c^{-2}\lambda\sin\bm{Q}\cdot\bm{\delta})^2/2\kappa$ with the nearest-neighbor coordination vector $\bm{\delta}$. 

(iii) In the case of corner-sharing network of magnetic and ligand ions, four-spin ring-exchange interaction arises from the strong-coupling expansion. Substituting Eq.~(\ref{eq:S_r:ab}) into the ring-exchange Hamiltonian aorund four sites $\bm{r}_i$ ($i=1,\cdots,4)$ in a $xy$-plane plaquette,
\begin{equation}
  J^{\text{ring}}\left(P_{\bm{r}_1,\bm{r}_2;\bm{r}_3,\bm{r}_4}+P_{\bm{r}_2,\bm{r}_3;\bm{r}_4,\bm{r}_1}-P_{\bm{r}_1,\bm{r}_3;\bm{r}_2,\bm{r}_4}\right)
  \label{eq:H:ring}
\end{equation}
with $P_{\bm{r},\bm{r}';\bm{r}'',\bm{r}'''}\equiv(\vec{S}_{\bm{r}}\cdot\vec{S}_{\bm{r}'})(\vec{S}_{\bm{r}''}\cdot\vec{S}_{\bm{r}'''})$ for spin-$1/2$~\cite{SchmidtKuramoto,Brehmer99}, we obtain $\delta u^{\text{ring}}=c^{-4}J^{\text{ring}}(\cos^2Q_x+\cos^2Q_y-1/2)$ and $\delta v^{\text{ring}}=c^{-4}J^{\text{ring}}(\sin^2Q_x+\sin^2Q_y-1/2)$. 

On the other hand, in the case of edge-sharing network, the above ring-exchange is suppressed, while the second-order strong-coupling perturbation of the four-site Coulomb interaction in the unit square gives a still fnite contribution of the form
\begin{equation}
  -K_1\left[P_{\bm{r}_1,\bm{r}_2;\bm{r}_3,\bm{r}_4}+P_{\bm{r}_2,\bm{r}_3;\bm{r}_4,\bm{r}_1}\right]-K_2 P_{\bm{r}_1,\bm{r}_3;\bm{r}_2,\bm{r}_4}
  \label{eq:H:ring2}
\end{equation}
where $K_1\sim 4I_{1234}^2/U$, $K_2\sim 4(I_{1234}-I_{1243})^2/U<K_1$, and $I_{ijk\ell}\equiv\int d\bm{\rho}d\bm{\rho}'\frac{e^2}{|\bm{\rho}-\bm{\rho}'|}\phi(\bm{\rho}-\bm{r}_i)\phi(\bm{\rho}'-\bm{r}_j)\phi(\bm{\rho}'-\bm{r}_k)\phi(\bm{\rho}-\bm{r}_\ell)$ with the local Coulomb repulsion $U$ and the Wannier function $\phi(\bm{\rho})$ centered at the magnetic ion site. Again substituting Eq.~(\ref{eq:S_r:ab}) into Eq.~(\ref{eq:H:ring2}), we obtain $\delta u^{\text{ring}}=-c^{-4}(2K_1+K_2)(\cos^2Q_x+\cos^2Q_y-1/2)$ and $\delta v^{\text{ring}}=c^{-4}K_1(\cos^2Q_x+\cos^2Q_y-1)+K_2/2$.

Putting the above considerations together, the deviation of the parameter $v/u$ from the Heisenberg value $-1/3$, which plays a crucial role in determining the phase diagram, is eventually given by
\[
  \frac{v}{u}+\frac{1}{3}\sim\sum_{\bm{\delta}}\frac{\sin^2\bm{Q}\cdot\bm{\delta}}{3c^4u_0}\biggl[\frac{\delta J^2}{2K}-\frac{3\lambda^2}{2\kappa}+4J^{\text{ring}}-(K_1-K_2)\biggr].
\]
Namely, the DM interaction and the Coulombic four-spin ring-exchange yield a negative shift of $u/v$ and favor the chiral spin pairing, whereas the magnetostriction and the kinetic four-spin ring-exchange interaction do not. 

Let us discuss a possibility of observing the spin cholesteric phase in real systems. First, the easy-axis single spin anisotropy is unfavorable, since it produces a mass between the two fields to be paired and suppresses the pairing. On the other hand, the chiral order remains in the easy-plane anisotropy since the chiral order parameter is now a scalar which forms a long-range order more easily than the O(3) vector.
In the case of $R$MnO$_3$ ($R$=Tb, Dy)\cite{Kimura03}, the spin
of the Mn ions is $S=2$ and the single spin anisotropy plays a dominant role
for the successive phase transitions from paramagnet to incommensurate collinear states and to incommensurate helical state as the temperature is lowered. Especially the transition to the helical state is of first-order, and hence the spin-pairing above the transition is unlikely. Therefore
$S=1/2$ is preferable. Then, the single spin anisotropy is forbidden, while the anisotropy of the exchange interaction is still possible. 
The edge-sharing network of magnetic and ligand ions is also advantageous over the corner-sharing, since it suppresses the magnetostriction and the kinetic four-spin ring-exchange interaction.
Cu ion with spin $S=1/2$ is an interesting candidate from this viewpoint. Actually multiferroic behaviors have been recently discovered in edge-sharing LiCuVO$_2$~\cite{LiCuVO2} and LiCu$_2$O$_2$~\cite{LiCu2O2}. Interestingly, the phase transitions to the helical state in both materials seem to be second order. A recent ESR experiment shows that the DM interaction of the order of 1 K survives~\cite{Mihaly06}. It is highly desirable to look for this intriguing possibility of chiral spin pairing by careful polarized neutron-scattering and ferroelectric polarization measurements slightly above the helical magnetic ordering temperature. 

The authors thank Y. Tokura, O. Entin-Wohlman, and A. Aharony for useful discussions and comments.
The work was supported by Grant-in-Aids under the Grant numbers 15104006, 16076205, and 17105002, and NAREGI Nanoscience Project from the Ministry of Education, Culture, Sports, Science, and Technology.

\end{document}